\documentclass[twocolumn,superscriptaddress,showpacs]{revtex4}
\usepackage{epsfig}
\usepackage{amsmath}
\usepackage{graphicx}

\begin{document}
\title{Classical simulability, entanglement breaking, and quantum computation thresholds}
\author {S. Virmani} \author {Susana F. Huelga}
\affiliation{School of Physics, Astronomy and Mathematics, Quantum
Physics Group, STRC, University of Hertfordshire, Hatfield, Herts
AL10 9AB, UK}
\author{Martin B. Plenio}
\affiliation{QOLS, Blackett Laboratory, Imperial College London,
London SW7 2BW, UK}

\date{\today}
\begin{abstract}
We investigate the amount of noise required to turn a universal
quantum gate set into one that can be efficiently modelled
classically. This question is useful for providing upper bounds on
fault tolerant thresholds, and for understanding the nature of the
quantum/classical computational transition. We refine some
previously known upper bounds using two different strategies. The
first one involves the introduction of bi-entangling operations, a
class of classically simulatable machines that can generate at
most bipartite entanglement. Using this class we show that it is
possible to sharpen previously obtained upper bounds in certain
cases. As an example, we show that under depolarizing noise on the
controlled-not gate, the previously known upper bound of 74\% can
be sharpened to around 67\%. Another interesting consequence is
that measurement based schemes cannot work using only 2-qubit
non-degenerate projections. In the second strand of the work we
utilize the Gottesman-Knill theorem on the classically efficient
simulation of Clifford group operations. The bounds attained using
this approach for the $\pi/8$-gate can be as low as 15\% for
general single gate noise, and 30\% for dephasing noise.

\end{abstract}


\maketitle
\section{Introduction}

The recent development of quantum information has led to a great
deal of interest in the classical simulation of quantum systems.
An understanding of this issue is important in order to discern
which resources are essential for an exponential quantum speedup.
If we remove certain resources from a particular model for
universal quantum computation, and find that the resulting machine
can be efficiently simulated classically, then we can infer that
those resources are essential to any exponential speedup that the
original device may offer. For instance, this approach has been
used to show that quantum entanglement is an essential ingredient
for quantum computation \cite{loads,AB,HN} while fermionic linear
optics does not allow for an exponential speedup \cite{V}.

In addition to questions of resources, an understanding of
classically tractable quantum evolution is also useful for
bounding the fault tolerance thresholds of universal quantum
machines. This connection becomes apparent from another important
question concerning any universal quantum machine: {\it what is
the minimal amount of noise required before the device can be
efficiently modelled classically?}. We loosely refer to this
minimal noise level as the {\it classical tolerance} of a
particular physical machine. We will also use the term {\it
tractable} to describe any form of quantum evolution that may be
modelled with polynomial {\it classical} resources. If it is true
that quantum computation is not tractable classically, then upper
bounds to the classical tolerance of the gates in a universal
quantum gate set are also upper bounds to the fault tolerance of
those gates. Aharonov and Ben-Or were among the first to obtain
upper bounds on the classical tolerance thresholds of quantum
gates \cite{AB}. To obtain their bounds they assumed that noise
acts on every qubit at every stage of the computation, and showed
that for noise above a certain amount the evolution becomes
classically tractable (see also \cite{HN,R} for related work).

In addition to bounding fault tolerance, there is perhaps a more
fundamental reason for investigating where the classical/quantum
computational transition lies \cite{AB,BK}. It may well be the
case that noisy quantum devices cannot be simulated efficiently
classically, yet cannot be used for fault tolerant quantum
computation. This would imply the existence of an `intermediate'
physical device - such as a noisy quantum system controlled by a
universal classical computer - which is clearly universal for
computation, is better-than-classical as it can simulate itself
efficiently, and yet is not as powerful as a full quantum
computer. Hence classical tolerance thresholds also provide
important (and perhaps easier) milestones for experimental
efforts.

In this work, however, we will be more interested in the recent
approach taken by Harrow and Nielsen \cite{HN} where they
presented an algorithm for the efficient classical simulation of a
quantum machine operating with {\it separability preserving}
quantum gate sets (`{\it SP machines}'). The term {\it
separability preserving} refers to any set of operations that
cannot entangle product inputs. They then derived bounds on the
minimal noise levels required to turn certain universal quantum
gate sets into SP machines, thereby obtaining bounds on the
classical tolerance of those gates. Due to the lack of a simple
characterization of the SP machines, in most cases their
calculations proceeded not by considering the full set of SP
machines, but instead the set of {\it separable machines}, which
are those devices that only operate with {\it separable} quantum
gates \cite{rains}. Their approach has the advantage that one can
even consider weak noise models where the noise only acts whenever
multi-qubit gates are applied. Depending upon the noise model,
however, the upper bounds to classical tolerance derived in this
way were of the order of 50\% or more for interesting universal
gate sets such as CNOT+single qubit operations. In terms of
depolarizing noise only, Razborov has obtained the strongest
bounds that we are aware of - showing that for two-qubit gates
50\% noise is an upper bound to fault tolerance \cite{R}. However,
his approach cannot be directly compared to that of \cite{HN}, as
it does not consider efficient classical simulation, and assumes a
different noise model (where each qubit is decohered at every
timestep).

In this article we will consider efficient classical simulation,
and we will extend the approach taken in \cite{HN} along two
different tracks. In the first track we define a class of quantum
machines that can generate entanglement between product input
states, but without additional resources can only generate at most
two-particle entanglement. We refer to any machines that operate
with our class of operations as {\it bi-entangling} machines, or
simply `{\it B-machines}'. A small extension of the algorithm
presented in \cite{HN} shows that such B-machines can be
efficiently simulated classically. We find that many of the
classical tolerance bounds derived in \cite{HN} are actually also
optimal with respect to B-machines. However, one example of an
improvement is the case of the CNOT under individual depolarizing
noise on the qubits, where we show that a 67\% noise rate leads to
classical tractability, which is stronger than the 74\% bound
derived in \cite{HN} for the same model. Another interesting
example comes from 2-qubit measurement based quantum computation,
where we find that exponential speedup requires degeneracy in at
least one of the projections - a result that cannot be directly
derived from the approach in \cite{HN}, and suggests that for
noise models in measurement based quantum computation our approach
could be more fruitful. As an aside we also observe that there are
separability preserving gates that are not probabilistic mixtures
of separable/ separable+swap operations, thereby deciding a
conjecture made in \cite{HN}.

In the second track we make use of the Gottesman-Knill theorem
\cite{GK}. All of the results discussed above are derived by
considering machines that create a limited amount of entanglement,
or are so noisy that they tend to some form of equilibrium.
However, the important {\it Gottesman-Knill} theorem states that
machines composed of Clifford group unitaries \cite{C} and
computational basis state preparation \& measurement {\it can} be
efficiently modelled classically, despite the fact that such
resources are capable of generating many-particle entanglement
(although not all forms of entanglement \cite{W}). It is hence
natural to ask whether such {\it Clifford machines} can lead to
better bounds on classical tolerance than bi-entangling or SP
machines. We calculate exactly the minimal noise required to take
a variety of single qubit gates into the set of {\it Clifford
operations} - those operations that may be implemented by Clifford
group unitaries, computational ancillae, and measurements in the
computational basis. For the $\pi/8$ gate in particular \cite{NC},
for generic single operation noise, the bound obtained is
approximately 15\%, thereby showing that the $\pi/8$ gate in the
standard universal set {$\{\pi/8$,Clifford unitaries$\}$} cannot
be made fault tolerant to more than 15\% general individual gate
noise. For dephasing noise the bound is approximately 30\% for the
$\pi/8$-gate.

This paper is structured as follows. In the next section we
discuss the class of bi-entangling machines and the reasons why
they can be efficiently modelled classically and discuss the
classical algorithm. In section III we discuss the way that we
will choose to represent quantum operations - via the {\it
Jamiolkowski isomorphism} \cite{J}, and derive some classical
tolerance bounds with respect to B-machines. In section IV we
derive bounds for Clifford operations based gate sets by using the
Gottesman-Knill theorem. In section V we discuss some subtleties
in the interpretation of results from section IV. Section VI is
the conclusion.

\section{The Bi-entangling machines}

We define the term {\it Bi-entangling machine} according to the
following:

\medskip

\noindent {\bf Definition:} A {\it Bi-entangling machine}
(`B-machine') is one that consists of a supply of individual
qubits initialised in some fixed state, augmented by the following
quantum operations:

\begin{enumerate}
\item an arbitrary set of single qubit quantum operations (these
may be unitary, or measuring, or anything else),

\item an arbitrary set of 2-qubit operations that can be
expressed as convex combinations of (a) separable operations
\cite{rains} that do not entangle the two qubits, (b) operations
that swap the two qubits and then apply a separable operation, and
(c) entanglement breaking (EB) \cite{EB} operations that break any
entanglement between the two qubits and the rest of the qubits.
\end{enumerate}

\medskip

The fact that any machine consisting of (1), (2a,b) is efficiently
classically tractable was already shown in \cite{HN}, as such
operations lie (strictly) within the set of separability
preserving operations. The only new point added here is the
inclusion of operations from (2c), and the resulting convex hull
with the separable/separable with swap operations. The heuristic
explanation for the algorithm is that a machine consisting of
operations (1)-(2) above only has the power to generate 2-particle
entanglement. The technical details of the proof then involve only
small modifications to the algorithm presented by Harrow and
Nielsen \cite{HN}.

Suppose that a quantum machine starts in an initial state with the
qubits in a product state, and that subsequent evolution is only
via bi-entangling operations. We will assume w.l.o.g. that there
are $4N$ qubits involved in the machine, where $N$ is an integer.
In order to model the system, we must do two things: (i) we must
represent the state of the machine, and (ii) we must track the
evolution of the system.

(i) Firstly, let us consider how we will represent the state. For
reasons that will become apparent later, we first partition the
qubits into 2N pairs, such that each qubit is allocated exactly
one partner. If all the qubits are initially unentangled, we may
make an arbitrary choice. We store a database of the pairs in the
memory of our classical device. As there are 2N pairs, the memory
is of the order of 2NlogN (the logN comes from the fact that we
need a label for each qubit). We refer to this database as the
{\it pairing list}. In addition to the pairing list, we must also
store the quantum state of the machine. This is done using 2N
vectors $\{{\bf r}_1,{\bf r}_2,..,{\bf r}_{2N}\}$, where each
${\bf r}_i$ is a 16-component vector that represents the (possibly
mixed) state of the 2-qubit couple $i$. We will restrict ourselves
to storing each ${\bf r}_i$ to some fixed accuracy, and so we only
require a memory proportional to 2N to store all of these vectors.
In principle it will be necessary for us to understand how this
limited accuracy will affect our simulation. This is a non-trivial
point, however, the approach developed in pages 3-6 of \cite{HN}
to tackle this problem also applies here, and so we do not discuss
the details.

(ii) We now need to understand how to track the evolution. We need
a way of representing each particular quantum gate on the
classical computer. The first gate that we apply may be either a
single qubit gate, or a 2-qubit gate that is a convex combination
of separable and entanglement breaking operations. We deal with
each case separately:

(a) A single qubit gate. Suppose w.l.o.g. that the first and
second qubits are in a pair, and that the gate is applied to the
first qubit. Each single qubit gate can be considered as a linear
transformation on one particle from a pair, and therefore
represented as a $16 \times 16$ matrix acting upon the vector
${\bf r}$ that describes the pair in question. To track the
changes due to single qubit gates, we hence simply need perform a
single linear transformation of a single vector ${\bf r}$ in our
memory.

(b) A two qubit gate. Suppose that we are implementing a two qubit
gate. There are two possibilities. The two input particles are
already in a pair, or they are in separate pairs. If they are
already in a pair, we merely update the vector ${\bf r}$ of the
pair according to the whole gate. If the two input particles are
not already in a pair, we first must throw a dice to decide
whether we will perform ($\alpha$) the separable part of the
evolution, ($\beta$) the separable+swap part of the evolution, or
($\gamma$) the entanglement breaking part.

($\alpha$) If we choose to implement the separable part, each
particle remains in the same pair, and we merely probabilistically
update the state of each pair.

($\beta$) If we choose the separable+swap part, we do the same
thing, except we first swap the particles in each pair according
to the swap operation.

($\gamma$) If we choose to implement the entanglement breaking
part, then the gate also enforces some partner-swapping, but not
quite so straightforwardly. Suppose that particles (1,2) and (3,4)
are in separate pairs, and that we use an EB gate to act upon
particles 2 and 3. To simulate its effects, we note that any EB
gate can be represented as a measurement followed by a state
preparation conditional upon the outcome. We hence calculate the
probability of the various outcomes corresponding to the
measurement, and with the correct probability follow one
particular outcome. Any particular outcome can result in the
pairing to be changed to (1,4) and (2,3), such that each new pair
has its own vector ${\bf r}$. So we both update the pairing list
to represent this fact, and compute the new vectors. Performing
this calculation corresponds to performing linear calculations
with vectors representing 4 qubits, and hence has a fixed upper
cost per EB gate being considered.

\medskip

We iterate these techniques for every gate that we apply, such
that at each stage we have a pairing list, and a set of vectors
representing the state of each pair. At the end of the whole
procedure we will need to simulate the outcomes of measurements on
individual qubits, which can be done using very similar methods to
those outlined above. Hence we can see that we have a classically
efficient algorithm for tracking the evolution of our device if
our multiparty gates are bi-entangling channels with fan-in $\leq
2$ (the {\it fan-in} of a gate is the number of particles that it
acts upon non-trivially).

One might hope that the algorithm may be extended, either to gates
with higher fan-in, or by incorporating all SP operations as well
as separable/ {separable+swap} operations. However, this cannot be
done straightforwardly. In the next section we will discuss why we
cannot include all SP operations.

To see why we cannot extend the fan-in of the gates either, it is
interesting to consider the connection between the above algorithm
and measurement based quantum computation schemes
\cite{cluster,leung,GC}. This situation also provides a simple
first example of where consideration of bi-entangling machines may
yield more information than consideration of SP machines alone. In
measurement based computation schemes it is known that two-qubit
measurements allow universal quantum computation \cite{leung}.
However, our algorithm shows that we must allow these measurements
to be degenerate, because if they are non-degenerate, then the
resulting operations will be EB, and the device cannot offer an
exponential speedup. This leads to a useful rule: {\it any 2-qubit
measurement based scheme for quantum computation must involve
non-degenerate measurements}.

Although this observation is quite simple, it applies to gates
that can generate {\it some} entanglement (e.g. Bell
measurements), and so it cannot be derived directly from the
approach in \cite{HN}. However, this limit on the capacity to
generate multi-particle entanglement is removed when you allow EB
channels with 3 or more inputs, and this is one reason why
universal quantum computation is possible using some forms of
non-degenerate measurements on 3 or more particles (see e.g. the
paper by Gottesman \& Chuang on teleportation based computation -
they use GHZ-like states and Bell measurements \cite{GC}).
Therefore it is difficult to extend the bi-entangling class to
gates acting on three or more parties.

\section{Representation of quantum operations by states}

In order to utilise the above algorithm to bound the classical
tolerance of quantum gates, it is important to be able to decide
when a given set of quantum operations falls into the class of
B-machines. In general this problem is extremely difficult.
However, some important operations such as the CNOT gate possess a
great deal of symmetry that makes the analysis tractable. In order
to perform this analysis, the {\it Jamiolkowski isomorphism}
\cite{J} provides a convenient way of representing quantum
operations. To any trace preserving quantum operation ${\cal E}$
on a single particle of $d$-levels, the Jamiolkowski isomorphism
associates a two-party quantum state that we will refer to as the
{\it Jamiolkowski state}, $\rho({{\cal E}})$:
\begin{equation}
\rho({{\cal E}}):= I_A \otimes {\cal E_B}(|+\rangle\langle+|)
\end{equation}
where $|+\rangle := (1/\sqrt{d})\sum_{i=1}^{d} |ii\rangle$ is the
canonical maximally entangled state for two {$d$-level} systems
$A,B$. It is clear from the above definition that $\rho({{\cal
E}})$ has a reduced density matrix $(\rho({{\cal E}}))_A$ that is
maximally mixed. It turns out that any density matrix with this
property (i.e. one with $\rho_A$ maximally mixed) can be
associated with a quantum operation ${{\cal E}}$. Moreover, a
simple teleportation like argument can be used to show that this
association is one-to-one. Hence the Jamiolkowski isomorphism is a
one-to-one mapping between the set of trace preserving quantum
operations ${\cal E}$ and two-party quantum states with maximally
mixed reduced density matrix.

This isomorphism can be easily applied to multi-party quantum
operations in the following way. Suppose that we have a two
particle quantum operation ${\cal E}_{12}$ acting upon two qubits
$1,2$. To represent this operation we must use a quantum state of
four parties $A1,A2,B1,B2$ \cite{cirac}:
\begin{eqnarray}
&& \rho({{\cal E}}_{1,2}):= \nonumber \\ && I_{A1} \otimes I_{A2}
\otimes {\cal E}_{B1,B2}\left( (|+\rangle\langle+|)_{A1,B1}
\otimes (|+\rangle\langle+|)_{A2,B2}\right) \nonumber
\end{eqnarray}
This representation is particularly convenient because various
important properties of quantum operations ${\cal E}$ may easily
be translated into properties of the corresponding state
$\rho({\cal E})$. In this work we will consider three such
properties (see figure (\ref{ops})):
\begin{figure}
 \begin{center}
  \includegraphics[scale=0.35]{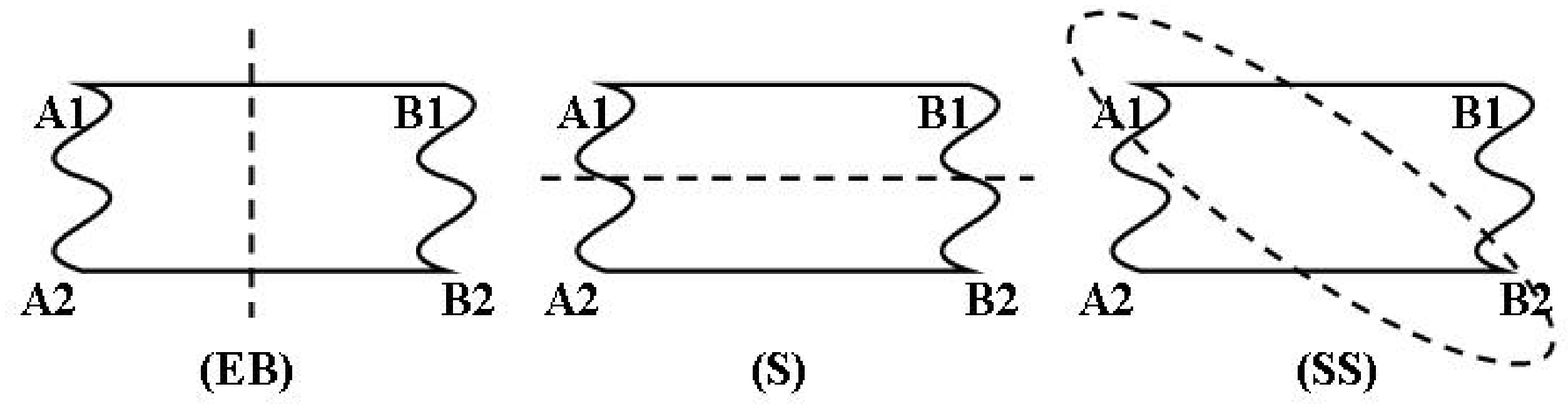}
  \caption{Entanglement breaking (EB), separable (S), and separable+swap (SS) operations have
  a simple connection to Jamiolkowski state separability. For EB and separable operations (S), the
  dashed line indicates the corresponding separable split. For separable+swap (SS) operations the dashed
  ellipse indicates the splitting. In all diagrams the wavy lines indicate entanglement between pairs 1 \& 2.}
 \label{ops}
 \end{center}
\end{figure}

\bigskip

\indent (a) an operation is separable iff the Jamiolkowski state
\indent is separable across the (A1A2)-(B1B2) split.

\medskip

\indent (b) an operation is equivalent to the swap operation,
\indent preceded by and followed by separable operations, iff
\indent the Jamiolkowski state is separable across the (A1B2)-
\indent (A2B1) split.

\medskip

\indent (c) an operation is entanglement breaking iff the \indent
Jamiolkowski state is separable across the (A1B1)- \indent (A2B2)
split.

\bigskip

A set of operations is bi-entangling if every operation lies
within the convex hull of these three classes (a),(b) and (c). An
operation is hence bi-entangling iff the Jamiolkowski state that
represents it can be written as:
\begin{eqnarray}
\rho = p \sum_i p_i \rho^i_{A1A2} \otimes \rho^i_{B1B2} + q \sum_j
q_j \rho^j_{A1B2} \otimes \rho^j_{A2B1} \nonumber \\ + r \sum_k
q_k \rho^k_{A1B1} \otimes \rho^k_{A2B2}
\end{eqnarray}
where $(p,q,r)$ is a probability distribution, the sets $\{p_i\}$,
$\{q_j\}$, $\{r_k\}$ are also individual probability
distributions, and all {$\rho$'s} on the r.h.s. are valid density
matrices.

This is a convenient point to discuss the relationship between
B-machines and SP-machines. It is clear from the above definition
that B-machines contain the convex hull of the separable
operations with separable+swap operations. However, they do not
contain all possible separability preserving operations. This
becomes apparent from consideration of the following Jamiolkowski
state:
\begin{eqnarray}
\omega : = && {1 \over 2}|GHZ\rangle \langle GHZ|_{A1,A2,B1}
\otimes |0\rangle \langle 0|_{B2} \nonumber \\ + && {1 \over 2}
|GHZ'\rangle \langle GHZ'|_{A1,A2,B1} \otimes |1\rangle \langle
1|_{B2}
\nonumber \\
\mbox{where:} &&|GHZ\rangle := {1 \over \sqrt{2}}(|000\rangle
+|111\rangle) \nonumber\\
&&|GHZ'\rangle := {1 \over \sqrt{2}}(|011\rangle +|100\rangle)
\end{eqnarray}
As $\omega$ is a valid density matrix with the reduced state of
parties A1-A2 maximally mixed, it corresponds to the Jamiolkowski
state of a valid quantum operation. When viewed as a state of four
parties, $\omega$ also has the property that a GHZ-type state can
be distilled from it by LOCC operations - simply by measuring the
particle B2 in the computational basis. However, as the
Jamiolkowski states of bi-entangling operations contain only
two-particle entanglement, this means that $\omega$ cannot
represent a bi-entangling operation. However, $\omega$ manifestly
represents an SP operation, because the output qubits B1-B2 are
always left in a separable state. Therefore we can also conclude
that the conjecture made in \cite{HN} that that {SP operations
$\neq$ convex hull$\{$separable ops, separable+swap$\}$}, is
indeed true \cite{foot}. In the above definition of B-machines, we
have included the ability to make separable operations,
separable+swap operations, and EB operations. One might be tempted
to expand this definition to include all SP operations as well.
However, operations such as $\omega$ have the capacity to
probabilistically generate many-particle entanglement when
accompanied by EB channels such as Bell-measurements, and so in
our definition of B-machines we are forced to include the smaller
classes of separable and separable+swap operations, and not the
larger class of SP operations.

We would now like to use the class of B-machines defined above to
obtain bounds on the classical tolerance of important universal
gate sets. Suppose for example that we have a universal quantum
computer consisting of the CNOT gate and a sufficient set of
single qubit operations. If we add some noise to the CNOT such
that it is taken to a bi-entangling operation, then the whole set
is taken to a B-machine, and can be efficiently classically
simulated. Hence to bound the classical threshold of the CNOT in
our device, we would like to calculate the minimal noise required
to turn the CNOT into a bi-entangling operation. In general such
calculations are very difficult.

It is at this point that we must discuss the form of the noise
model that we consider. In the rest of this article we adopt the
standard {\it probabilistic} noise model, where qubits are
affected incoherently. In this model whenever we would like to
perform an ideal quantum operation ${\cal E}$, instead due to
noise we are forced to perform an operation ${\cal E'}$ that is
related to ${\cal E}$ as follows:
\begin{equation}
{\cal E'} = (1-p){\cal E} + p{\cal N} \label{model}
\end{equation}
where $p$ is a probability, and ${\cal N}$ is some other quantum
operation that represents the error. In this equation $p$ is
measure of the error rate. Note that this is not the most general
model of error, and not necessarily the most physical model
either. Consider the example where our ideal operation is to
simply preserve the state of a qubit, but in fact it undergoes a
spontaneous emission at a sufficiently slow rate. This form of
error cannot be written in the form of equation (\ref{model})
unless the error parameter is set to $p=1$ (see e.g. \cite{NC},
page 442). For more generic errors one would have to adopt some
suitable metric $\|\bullet\|$ on the set of quantum operations and
use $\|{\cal E'}-{\cal E}\|$ as a measure of error rate (see e.g.
\cite{GLN} for some possible metrics). Although several authors
have considered more general models of error in relation to fault
tolerance \cite{noise}, the only prior work on classical tolerance
has been within the framework of equation (\ref{model}), and this
is the model that we will follow here. In the case of Markovian,
identical, and independent noise it should be possible to extend
many of the techniques presented here to metric based noise
quantification, although we will not pursue that avenue here.
Within the probabilistic model, one can also make further
restrictions, and constrain the form of ${\cal N}$ to interesting
forms of noise such as depolarization or dephasing. As with
\cite{HN}, however, our analysis will initially take ${\cal N}$ to
be a general quantum operation.

Given that our error model is probabilistic, our task is to find
the minimal value $p$ such that there is a valid ${\cal N}$ taking
our ideal gate into the set of bi-entangling operations. In
general this is likely to be a difficult task. However, for the
case of the CNOT a great deal of symmetry is present that enables
the calculation to be performed exactly. In order to see how this
proceeds, it will be first helpful to consider the case that the
two qubit gate is a general unitary $U$, and examine some of the
symmetry possessed by the Jamiolkowski state that represents $U$.

For any 2-qubit unitary $U$ we have the trivial identity:
\begin{equation}
(U(\sigma_i \otimes \sigma_j)U^\dag )U (\sigma_i \otimes \sigma_j)
= U
\end{equation}
where $\{\sigma_i|i=0,x,y,z\}$ are the standard Pauli operators.
This identity, together with the fact that $I \otimes A |+\rangle
= A^T \otimes I |+\rangle$ for any linear operator $A$, can be
used to show that the Jamiolkowski state representing $U$ commutes
with all operators of the form:
\begin{equation}
W_{ij}^U:=(\sigma_i^T)_{A1} \otimes (\sigma_j^T)_{A2} \otimes
(U(\sigma_i \otimes \sigma_j)U^\dag )_{B1,B2}. \label{commutant}
\end{equation}
It is not hard to verify that as we vary over $i,j$
the operators in equation (\ref{commutant}) form a group (up to an
unimportant phase), and moreover
 from the commutation
relationships of the Pauli operators it follows that the group is
abelian. It hence follows from Schur's lemma that any operator
that commutes with all operators of the form (\ref{commutant}) is
diagonal in the eigenbasis formed by the one-dimensional
irreducible representations of the group (\ref{commutant}). We can
construct these irreducible representations quite easily. In fact,
the group (\ref{commutant}) is isomorphic to the group consisting
of elements
\begin{equation}
(\sigma_i^T)_{A1} \otimes (\sigma_j^T)_{A2} \otimes
(\sigma_i)_{B1} \otimes (\sigma_j)_{B2},
\end{equation}
as it is related to (\ref{commutant}) by the unitary
transformation $I_{A1} \otimes I_{A2} \otimes U_{B1,B2}$. Hence we
can utilize the stabilizer formalism for the Pauli group, and
write the 16 common eigenstates of the operators in
(\ref{commutant}) as:
\begin{eqnarray}
|e,U\rangle\langle e,U|:=  &&\left( {I + (-1)^{e_0}W^U_{0x}\over
2}\right)
 \left( {I + (-1)^{e_1}W^U_{0z}\over 2}\right) \times \nonumber \\
&& \left( {I + (-1)^{e_2}W^U_{x0}\over 2}\right)\left( {I +
(-1)^{e_3}W^U_{z0}\over 2}\right)
\end{eqnarray}
where $e$ is a 4-bit string given by its components $e_{\alpha}
\in \{0,1\}, \alpha=0,1,2,3$. It turns out that each of these
eigen-projectors $|e\rangle\langle e|$ is a Jamiolkowski state for
a valid quantum operation - the normalization and positivity are
automatic, and the reduced density matrices over particles A1,A2
are all maximally mixed (this is in turn because $\sigma_i \otimes
\sigma_j$ is an irreducible representation). The Jamiolkowski
state representing $U$ is in fact given by the projector
$|e=0,U\rangle\langle e=0,U|$ corresponding to $e=0$:
\begin{equation}
\left( {I + W^U_{0x}\over 2}\right)
 \left( {I + W^U_{0z}\over 2}\right)
\left( {I + W^U_{x0}\over 2}\right)\left( {I + W^U_{z0}\over
2}\right)
\end{equation}
If we denote the Jamiolkowski state that represents $U$ by
$\rho(U)=|e=0,U\rangle\langle e=0,U|$, then our task is to find
the minimal probability $p$ such that for some quantum noise
${\cal N}$:
\begin{equation}
{\cal{E}} = (1-p)\rho(U)+ p\rho({\cal N}) \label{task}
\end{equation}
is the Jamiolkowski state of a bi-entangling operation. Now the
properties of the CNOT allow us to make further simplifications.
The CNOT is a member of the Clifford group, meaning that for any
two Pauli operators $\sigma_i, \sigma_j$ we have that
\begin{equation}
CNOT(\sigma_i \otimes \sigma_j)CNOT \sim \sigma_k \otimes \sigma_l
\end{equation}
where $\sigma_k, \sigma_l$ are other Pauli operators, and the
symbol $\sim$ means that the two sides of the equation are equal
up to an unimportant global phase. This means that the group
(\ref{commutant}) corresponding to the CNOT is actually a local
group, where each element is a tensor product of Pauli operators
acting on individual qubits of the Jamiolkowski state.
 We can therefore
average (`twirl') over the group (\ref{commutant}) any valid
solution (\ref{task}) corresponding to the CNOT, and as each
$W^{CNOT}_{ij}$ is local, the bi-entangling properties of the
equation will not be changed. This means that without loss of
generality, for the CNOT we need only consider `twirled' noise
states {{${\rho}'$}$({\cal N})$} that are also invariant under the
action of the group. This means that we can set
\begin{equation}
\mbox{{{${\rho}'$}$({\cal N})$}} = \sum_e \lambda_e({\cal N})
|e\rangle\langle e|
\end{equation}
where $\{\lambda_e\}$ is a probability distribution of
eigenvalues. If we have not constrained further the form of ${\cal
N}$, then the form of the probability distribution
$\{\lambda_e({\cal N})\}$
 can be left free. However, if we are restricting ${\cal N}$ to be of a specific
 form such as depolarization or dephasing, then we will have to restrict
 the distribution accordingly. Our task is hence to find the minimal probability $p$ such that
there exists a probability distribution $\{\lambda_e({\cal N})\}$
(consistent with any further constraints upon the noise) such that
the state:
\begin{eqnarray}
(1-p)\rho(U) + p\left(\sum_e \lambda_e ({\cal N}) |e\rangle\langle
e|\right)
\equiv \nonumber \\
 \left((1-p)+p\lambda_0 ({\cal N})\right) |e=0\rangle\langle
e=0| + \sum_{e\neq 0} \lambda_e ({\cal N}) |e\rangle\langle e|
\label{optimization}
\end{eqnarray}
is bi-entangling. Let us denote this optimal value of $p$ by
$p_{min}$. We can now try to perform this optimization for various
possible constraints upon the noise:

\bigskip

\noindent(a) {\bf No constraints:} In this case we need only
restrict the {$\lambda_e$s} to be a probability distribution, and
do not need to further constrain them. Take $\lambda^{opt}_0(B)$
to be the maximal possible $\lambda_0$ over all bi-entangling
states invariant under the symmetry group (\ref{commutant}). Then
we clearly have that:
\begin{equation}
(1-p) + p\lambda_0({\cal N}) \leq \lambda^{opt}_0(B)
\end{equation}
and hence as $p,\lambda_0 \geq 0$ we have that:
\begin{equation}
(1-p) \leq \lambda^{opt}_0(B) \Rightarrow p \geq
1-\lambda^{opt}_0(B)
\end{equation}
This lower bound can be attained as we are free to choose the form
of the noise as we wish. Hence $p_{min} = 1 - \lambda^{opt}_0$,
and our task is now to calculate $\lambda^{opt}_0$. This is now an
easier problem, as the fact that the set of bi-entangling states
is the convex hull of separable, separable+swap, and EB states
means that:
\begin{equation}
\lambda^{opt}_0(B) =
\mbox{max}\{\lambda^{opt}_0(S),\lambda^{opt}_0(SS),\lambda^{opt}_0(EB)\}
\end{equation}
where $\lambda^{opt}_0(S),\lambda^{opt}_0(SS),\lambda^{opt}_0(EB)$
are the maximal possible {$\lambda_0$'s} over separable states,
separable+swap states, and EB states respectively. This means that
to work out the minimal generic noise required to turn the CNOT
into a bi-entangling gate, we simply need to separately calculate
the minimal noise required to take the CNOT into the different
classes of separable, separable+swap, and EB, and take the lowest
value. As each of these classes separately corresponds to
separability across a particular partition of the parties in the
Jamiolkowski state, we can apply the techniques developed in
\cite{HN}. Although we omit the details, it turns out that the
Jamiolkowksi state representing the CNOT has only 1 ebit of
maximal entanglement across the (A1B1)-(A2B2) splitting or the
(A1B2)-(A2B1) splitting, but as with any two-qubit unitary has a
full 2 ebits of maximal entanglement across the (A1A2)-(B1B2)
splitting. Hence the CNOT is less robust to noise across the
separable/separable+swap splittings. Furthermore, the results of
\cite{vidal,HN} show that the minimal noise that breaks the
entanglement of the CNOT across the relevant splitting can always
be chosen to be separable across that splitting. The result of all
these observations is that the bounds derived in \cite{HN} for
generic noise are also optimal when considering B-machines as the
classically tractable set. It is also interesting to note that if
we do not ask for the noise to take us into the bi-entangling set,
but instead ask to be taken into the entanglement breaking
channels, then the above approach yields solutions for {\it all}
unitary gates, not just the CNOT \cite{foot2}.

\bigskip

\noindent (b) {\bf Separable noise, Separable+swap noise, Noise
that is a mixture of Separable \& Separable+swap:} The previous
paragraph points out that by the arguments of \cite{vidal,HN}, the
minimal generic noise that turns the CNOT into a bi-entangling
gate can always be taken to be separable, or separable+swap. Hence
the bounds derived in \cite{HN} are also optimal w.r.t. these
forms of noise, and where the classically tractable set is the set
of B-machines.

\bigskip

\noindent (c) {\bf Depolarizing noise:} The case of depolarizing
noise on individual qubits is a little more tricky to handle than
two particle noise. This is primarily because the depolarization
is assumed to act independently upon the two qubits in the quantum
gate. If we adopt the model in equation (56) of \cite{HN}, with an
error parameter $p$ the noisy operation is in fact:
\begin{equation}
(1-p)^2 U + p(1-p)(D \otimes I)U + p(1-p)(I \otimes D)U + p^2 D
\otimes D \label{dep}
\end{equation}
where $D$ represents the single qubit depolarizing quantum
operation,
\begin{equation}
D: \rho \rightarrow {I\over 2},
\end{equation}
and $U$ represents the ideal unitary quantum operation (we often
represent a unitary and the corresponding quantum operation by the
same letter - the meaning should be clear from the context). In
particular we will take $U$ to be the CNOT operation. In order to
derive un upper bound on the minimum value of $p$ required to make
this noisy operation bi-entangling, we will first show that the
(unnormalised) quantum operation corresponding to the central
terms of equation (\ref{dep}):
\begin{equation}
p(1-p)[(D \otimes I)U + (I \otimes D)U] \label{central}
\end{equation}
is in fact a separable operation (not just SP) for any value of
$p$. Hence if the (unnormalised) operation corresponding to the
outer terms:
\begin{equation}
(1-p)^2 U + p^2 D \otimes D \label{outer}
\end{equation}
is entanglement breaking, then the whole operation (\ref{dep}) is
bi-entangling. First we must show that the central terms
(\ref{central}) correspond to a separable operation. Consider the
operation $(D \otimes I)U$, where $U$ is the CNOT. After a little
algebraic manipulation of the Jamiolkowski state corresponding to
the CNOT, it can be shown that the Jamiolkowski state of $(D
\otimes I)U$ is:
\begin{eqnarray}
\rho((D \otimes I)U) = \left( {I + I_{A1} \otimes I_{B1} \otimes
X_{A2} \otimes
X_{B2}\over 2}\right) \times \nonumber\\
 \left( {I + Z_{A1} \otimes I_{B1} \otimes Z_{A2} \otimes
Z_{B2}\over 2}\right). \nonumber
\end{eqnarray}
Writing this out in the computational basis where $|0\rangle$
represents the +1 eigenstate of the $Z$ operator, and $|1\rangle$
represents the -1 eigenstate of the $Z$ operator, we find that
$\rho((D \otimes I)U)$ may be written as an equal mixture of the
following four pure states:
\begin{eqnarray}
(|0_{A1}0_{B1}\rangle \otimes \left({1 \over
\sqrt{2}}\right)(|0_{A2}0_{B2}\rangle+|1_{A2}1_{B2}\rangle),
\nonumber \\
(|0_{A1}1_{B1}\rangle \otimes \left({1 \over
\sqrt{2}}\right)(|0_{A2}0_{B2}\rangle+|1_{A2}1_{B2}\rangle),
\nonumber \\
(|1_{A1}0_{B1}\rangle \otimes \left({1 \over
\sqrt{2}}\right)(|0_{A2}1_{B2}\rangle+|1_{A2}0_{B2}\rangle),
\nonumber \\
(|1_{A1}1_{B1}\rangle \otimes \left({1 \over
\sqrt{2}}\right)(|0_{A2}1_{B2}\rangle+|1_{A2}0_{B2}\rangle).
\nonumber
\end{eqnarray}
As each of these pure states is separable across the (A1B1)-(A2B2)
split, it is clear that $(D \otimes I)U$ is a separable operation.
Similarly, one can show that the Jamiolkowski state representing
the operation $(I \otimes D)U$ is related to the state
representing $(D \otimes I)U$ in the following way:
\begin{equation}
\rho((I \otimes D)U) = \mbox{SWAP}_{1 \leftrightarrow 2}[
H^{\otimes 4} \rho((D \otimes I)U) H^{\otimes 4}]
\end{equation}
where the $H^{\otimes 4}$ is a Hadamard rotation on each qubit,
and {SWAP$_{1 \leftrightarrow 2}$} is the operation that
interchanges A1 with A2 and B1 with B2. As $\rho((I \otimes D)U)$
is related to $\rho((D \otimes I)U)$ by local rotations followed
by interchanging the labels $1 \leftrightarrow 2$, it is also
separable across the (A1B1)-(A2B2) split, and hence both central
terms in equation (\ref{dep}) correspond to separable operations.

It now remains for us to determine values of $p$ for which the
outer terms (\ref{outer}) represent an entanglement breaking
operation. The CNOT, as with any unitary on two-qubits, is
represented by a Jamiolkowski state that is maximally entangled
across the (A1B1)-(A2B2) splitting. The depolarizing operation on
both qubits $D\otimes D$, on the other hand, is represented by a
maximally mixed state. Hence if we are only considering the
(A1B1)-(A2B2) splitting, the state representing the operation of
equation (\ref{outer}) is essentially a maximally entangled state
of two 4-level systems, mixed with a maximally mixed state. The
conditions for such a state to be separable across the
(A1B1)-(A2B2) splitting, and hence entanglement breaking, are well
known, and correspond to:
\begin{equation}
{(1-p)^2 + p^2/16\over (1-p)^2 +p^2} \leq {1\over 4}
\end{equation}
giving that (\ref{outer}) is entanglement breaking whenever:
\begin{equation}
p \geq 2/3 \simeq 67\% .
\end{equation}
This means that the noisy CNOT gate is definitely bi-entangling
whenever the depolarizing noise rate is greater than 67\%. This is
an improvement over the 74\% bound derived in \cite{HN} for
exactly the same noise model, and hence shows that consideration
of B-machines may lead to tighter bounds than consideration of
separable machines alone. Of course the calculation here is not a
full optimization over all bi-entangling gates - we have only
calculated the minimal $p$ required to make the inner terms
separable, and the outer terms entanglement breaking. This hence
only provides an upper bound to the minimal $p$ required to make
the CNOT bi-entangling, and hence there is a possibility that this
calculation may be improved. However, as such full optimization is
likely to be difficult, we leave it to another occasion.

It is also worth noting that the classical tolerance bound of 67\%
derived here applies to any two-qubit unitary $W$ for which the
operations $(I \otimes D)W$ and $(D \otimes I)W$ are separable.

\section{Bounds from the Gottesman-Knill theorem}

In order to apply the Gottesman-Knill theorem \cite{GK} to
calculate bounds on the classical tolerance of quantum gates, we
need to compute the minimal amounts of noise required to take all
the gates in a particular machine into the Clifford class.
Unfortunately this restricts severely the possible situations in
which this approach may be applied. In previous examples we have
calculated the classical tolerance of certain two-qubit gates with
only very loose constraints on the other gates available to the
machine. In this section we will calculate the classical tolerance
of single qubit gates, assuming that the other gates in the
machine are Clifford operations, where we define Clifford
operations as follows:

\medskip

\noindent {\bf Definition:} {\it Clifford operations}. Those
operations that can be performed by probabilistic application of
Clifford group unitaries \cite{C}, (ancilla) state preparation,
and measurement in the computational basis.

\medskip

\noindent We will ask how much noise is required to turn
non-Clifford single qubit gates into a Clifford operation. The
resulting bounds on the classical tolerance can be relatively low.
For general single qubit operations we will show that the
classical tolerance to generic noise is no greater than 75\%,
although on a case-by-case basis this can be made much stronger.
For example for the $\pi/8$ gate, we find that 15\% noise is
minimal amount required to turn the gate into a Clifford
operation.

In order to perform these calculations, at first it seems
necessary to understand which single qubit operations can be
implemented using Clifford group unitaries and ancillas prepared
in the computational basis. However, we will not characterize this
set exactly here, as to obtain optimal bounds for many interesting
cases it turns out that it is sufficient to consider the effect
that Clifford operations have upon a particular subset of single
qubit states.

\begin{figure}
 \begin{center}
  \hspace{-1cm} \includegraphics[scale=0.4]{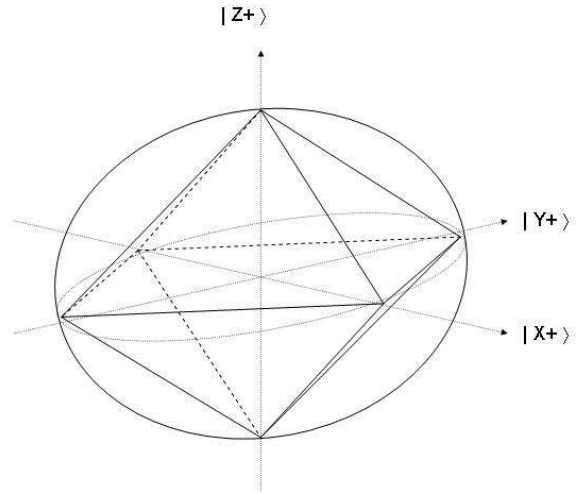}
  \caption{The accesible states via Clifford unitaries defines a octahedron in the Bloch sphere. Note that the vertices of the octahedron correspond
  to the Pauli eigenstates.}
 \label{octa}
 \end{center}
\end{figure}

We will consider the set of states that is given by the convex
hull of the Pauli operator eigenstates. This set is an octahedron
$O$ that is shown in figure (\ref{octa}). Our choice of this set
is inspired by the recent work of Bravyi \& Kitaev \cite{BK}, who
consider which single qubit state supplies may allow the Clifford
operations to become universal. We will first argue that the
octahedron $O$ can only be mapped to within itself by Clifford
operations, and use this fact to simplify the optimizations that
we wish to perform.

\smallskip

\noindent {\bf Observation 0:} The octahedron $O$ is closed under
the action of Clifford operations.

\smallskip

\noindent {\it Proof:} Let us consider a system $s$ that is
prepared in one of $\{|x\pm\rangle\langle
x\pm|,|y\pm\rangle\langle y\pm|,|z\pm\rangle\langle z\pm|\}$,
where $|a\pm\rangle$ refers to the up/down eigenstates of the
corresponding Pauli operator $A$. These states correspond to the
vertices of the octahedron $O$. Suppose also that there are $n-1$
ancillae prepared in the computational basis, as can be prepared
by Clifford operations. We need to calculate what possible final
states of the system are possible given Clifford group unitary
evolution of the system+ancilla \& Clifford group measurements. As
the entire input state is a stabilizer state \cite{GK,NC}, the
final state of system+ancilla will also be a stabilizer state that
is uniquely specified by its stabilizer generators
\begin{equation}
\{g_1,g_2 ..g_n\} \nonumber
\end{equation}
where each $g_i$ is a product of Pauli operators. Hence from the
standard theory of stabilizers, the final state of system+ancilla
will be given by:
\begin{equation}
\left({1 \over 2^{n}}\right)\prod_{i=1..n}(I + g_i) \nonumber
\label{states}
\end{equation}
This equation may be expanded, and {\it each element of the group
that is generated by the stabilizer will contribute exactly one
term in this expansion} (this follows from the independence of the
stabilizer generators \cite{NC}). As any non-trivial Pauli
operator is traceless, tracing out the $n-1$ ancilla qubits from
each term will only lead to a contribution to the final reduced
state of the system if the term is of the form $({1\over
2^{n}})A_s$, where $A_s := A_{\mbox{system}} \otimes I \otimes I
\otimes I \otimes... $, in which case the term will contribute
$A/2$ to the system density matrix. Our goal is hence to find
every group element of the form $A_s $ in the stabilizer group. As
the identity $I$ is an element in each stabilizer group, we will
at least have a contribution of $I/2$ (which is of course a
requirement in the Bloch expansion of any single qubit state).
However, we need to find all other terms of the form $A_s$.

This task can be constrained as follows. Firstly, in each
stabilizer group each element is its own inverse. This means that
any non-trivial terms of the required form must actually be one of
the 6 possibilities $\pm X_s, \pm Y_s$ or $\pm Z_s$. Moreover, at
most only {\it one} of these 6 possibilities is present in each
stabilizer group, as if two or more are present, then repeated
multiplication we would force $-I$ to be a member of the
stabilizer group (e.g. $(X_sY_s)^2=-I$), and this is not possible.
This means that input system states taken from the vertices of the
octahedron will be taken either to the maximally mixed state
$I/2$, or one of the eigenstates of the $X,Y,X$ operators
(corresponding to $(I/2 \pm X/2), (I/2 \pm Y/2)...$ etc.). This
means that the vertices will be taken either to the maximally
mixed state, or to another vertex. Then by convexity the
octahedron $O$ can only be mapped {\it onto or within} itself by
Clifford operations. \fbox \\

\medskip

\begin{figure}
 \begin{center}
  \hspace{-1cm} \includegraphics[scale=0.4]{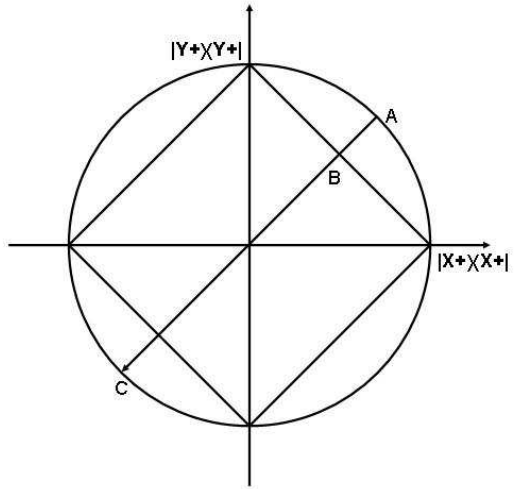}
  \caption{Cross section of the Bloch sphere in the x-y plane. Point
  A represents the state $|\psi(\pi/4)\rangle$, and the ratio {$|$AB$|$/$|$AC$|$} represents the
  exact minimal possible noise level required to take the $\pi/8$ gate into the set of Clifford operations.}
 \label{cool_fig}
 \end{center}
\end{figure}

This observation may be used to give {\it lower} bounds on the
amount of noise required to take any particular unitary operation
into a Clifford operation. Then by explicit construction we will
be able to show that whenever the unitary is diagonal in the
computational basis, that these lower bounds may be achieved, and
are hence tight. First let us see why the above arguments allow us
to construct lower bounds on the minimal noise level required.
Consider a unitary gate of the form:
\begin{equation}
U(\theta):=|0\rangle\langle 0| + \exp(i \theta) |1\rangle\langle
1|.
\end{equation}
This gate acts upon the $|x+\rangle$ state to give:
\begin{equation}
|\psi(\theta)\rangle := {1 \over \sqrt{2}}(|0\rangle + \exp(i
\theta) |1\rangle)
\end{equation}
We may visualize this by looking at the cross-section of the Bloch
sphere given by the x-y plane. This is shown in figure
(\ref{cool_fig}), with the point $A$ representing
$|\psi(\pi/4)\rangle$ corresponding to the action of the $\pi/8$
gate. One can see intuitively from the figure, and this can easily
be shown rigorously, that the minimal noise level required to take
the state $|\psi(\pi/4)\rangle$ into the octahedron is given by
the ratio {$|$AB$|$/$|$AC$|$} from the figure.
In the case of the $\pi/8$ gate, the ratio {$|$AB$|$/$|$AC$|$}
corresponds to a noise level of:
\begin{equation}
p = { \sqrt{2} - 1 \over 2 \sqrt{2}  } = 0.1464 \label{value1}
\end{equation}
If a noise level less than this amount could be added to the gate
$U(\theta)$ to turn it into a Clifford operation, then this would
mean that the $|x+\rangle$ state would be mapped to outside the
octahedron $O$ by the noisy operation. As this is not possible, we
can assert that (\ref{value1}) is a {\it lower} bound on the
amount of noise required to take the operation $U(\theta)$ into
the Clifford operations.

The utility of pictures such as figure (\ref{cool_fig}) is that
they may be used to show that bounds such as (\ref{value1}) are in
fact also upper bounds, and are hence tight. The argument for this
is strongly related to the construction presented \cite{BK} for
the programming of unitary operations in quantum states. Every
state on the circumference of the Bloch sphere in the x-y plane
corresponds to a pure state of the form:
\begin{equation}
|\psi(\theta)\rangle:={1 \over \sqrt{2}}(|0\rangle
+\exp(i\theta)|1\rangle)
\end{equation}
These states are clearly isomorphic to the Jamiolkowski states
representing each $U(\theta)$, simply by changing $|0\rangle
\rightarrow |00\rangle$ and $|1\rangle \rightarrow |11\rangle$:
\begin{equation}
|J(\theta)\rangle:={1 \over \sqrt{2}}(|00\rangle
+\exp(i\theta)|11\rangle).
\end{equation}
Hence by using convexity every state in the x-y cross section of
the Bloch sphere represents a valid quantum operation. From this
isomorphism we see hence that each of the vertices represents a
Clifford unitary: $|x+\rangle$ represents the identity gate
$\sigma_0$, $|x-\rangle$ represents the Pauli Z rotation,
$|y+\rangle$ state represents the so called {\it phase gate}
\cite{NC}, denoted by the letter $S$:
\begin{eqnarray}
  S := \left(\begin{array}{cc}
    1 & 0 \\
    0 & i \end{array}\right),
\end{eqnarray}
and $|y-\rangle$ represents its inverse, $S^{-1}$. This mapping
hence shows that bounds such as (\ref{value1}) can indeed be
attained, as the x-y plane of the Bloch sphere maps directly into
a problem concerning quantum operations and a subset of the
Clifford operations. Hence for gates of the form $U(\theta)$ we
have the following statement:

\medskip

\noindent {\bf Lemma 1:} The minimal noise required to turn
$U(\theta)$ into a Clifford operation is equivalent to the minimal
noise required to take the state $|\psi(\theta)\rangle$ into the
octahedron $O$ in figure (\ref{cool_fig}).

\medskip

As shown above, in the case of the $\pi/8$ gate this lemma returns
a minimal noise level is approximately 15\%. The same procedure
also yields sharp bounds for any unitary gate that may be
diagonalised by Clifford group unitaries, as well as for any
quantum operation that is a convex mixture of such unitaries.

We may also apply the above arguments to some cases where the
noise is constrained to be of a specific form. Suppose for example
that we wish to know how much dephasing noise is required to take
the $\pi/8$ gate into the set of Clifford operations. The
dephasing operation takes $|x+\rangle$ to the maximally mixed
state, at the centre of the Bloch sphere in figure
(\ref{cool_fig}). On the other hand, when figure (\ref{cool_fig})
is viewed as representing Jamiolkowski states of quantum
operations, the centre of the circle in figure (\ref{cool_fig})
also represents the dephasing operation. Hence the above arguments
also show that:

\medskip

\noindent {\bf Lemma 2:} The minimal dephasing noise required to
take any gate $U(\theta)$ into a Clifford operation is identical
to the minimal amount of maximally mixed state required to take
the corresponding state $|\psi(\theta)\rangle$ into the octahedron
$O$ in figure (\ref{cool_fig}).

\medskip

In the case of the $\pi/8$ gate this shows that approximately 30\%
dephasing noise is required to take the $\pi/8$ gate into the
Clifford operations, or more precisely twice the value in equation
(\ref{value1}):
\begin{equation}
 {\sqrt{2} - 1 \over  \sqrt{2}  } = 0.2928
\end{equation}
Although the bounds on the classical noise threshold obtained in
this way are quite low compared to bounds obtained in references
\cite{AB,HN,R}, the above procedure has the disadvantage that it
applies only to very specific gate sets, whereas previous works
have applied to much wider classes of machine. At the expense of
increasing the bound, we can however, make the approach more
general. For instance, we can show that the universal gate set
consisting of Clifford operations augmented by {\it any}
trace-preserving single qubit operation have a classical noise
tolerance of no greater than 75 \% on the additional single qubit
operation. The argument proceeds as follows. Given any
single-qubit trace preserving operation ${\cal{E}}$, we can always
turn it into an operation that is a convex mixture of Clifford
group operations by the following method. Instead of performing
${\cal{E}}$ on an input state $\rho$, we perform:
\begin{equation}
{1\over 4}{\cal{E}}(\rho)+ {3 \over 4} \sum_{i=x,y,z} {1 \over 3}
\sigma_i ({\cal{E}}(\sigma_i^T \rho \sigma_i^*))\sigma_i^\dag
\end{equation}
In the Jamiolkowski representation this quantum operation can be
represented as:
\begin{eqnarray}
&&{1\over 4} ( R_{\cal{E}} + ((\sigma_x \otimes \sigma_x)
R_{\cal{E}} (\sigma_x \otimes \sigma_x)^\dag)+ \nonumber
\\ && ((\sigma_y \otimes \sigma_y) R_{\cal{E}} (\sigma_y \otimes
\sigma_y)^\dag)+((\sigma_z \otimes \sigma_z) R_{\cal{E}} (\sigma_z
\otimes \sigma_z)^\dag)) \nonumber
\end{eqnarray}
This corresponds to a `Bell twirling', and the resultant quantum
operation is represented by a Bell diagonal state, which is a
mixture of the four Pauli transformations. Hence by adding 75 \%
noise, {\it any} trace preserving single qubit operations may be
taken to a probabilistic mixture of Clifford group operations, and
so any machine consisting of {\{CNOT+single qubit gates\}} has a
classical noise tolerance of at most 75 \% on the single qubit
gates.

\section{interpretation of the bounds.}

The bounds derived in the previous sections give upper bounds to
the fault tolerance of specific gates. For example, in the case of
the gate set {$\{$Clifford unitaries$, \pi/8-$gate$\}$}, they show
that no fault-tolerant encoding can be found that protects against
15\% {\it general} single gate noise. However, this does not mean
that specific forms of noise cannot be tolerated to greater than
15\%, but one must construct protection methods that specifically
target that form of noise.

Furthermore, in the case of the approach based upon the Clifford
group, our results show that if the Clifford gates in a gate set
are {\it noiseless}, then the noise corresponding to lemma 1 may
not be tolerated on an additional non-Clifford gate $U(\theta)$.
However, it is possible that by mixing noise in with the Clifford
gates as well, and not imposing that they be {\it noiseless}, one
can recover the power to do universal quantum computation. Indeed,
we have been able to construct examples where mixing a certain
type of noise to the gate $U$ from a universal set {$\{$Clifford
unitaries, $U\}$} leads to classically tractable evolution, but
mixing the same noise \cite{same} with the Clifford gates as well
as $U$ restores the ability to perform universal quantum
computation. Although we do not include the details here, the
examples that we have are all quite extreme, and work because
noise that turns the non-Clifford gate $U$ into a Clifford
operation can also take the Clifford unitaries {\it out} of the
Clifford group \cite{dep}. Nevertheless, in these examples the
fault tolerant encoding methods that restore universality are very
specific, and cannot be used to tackle general noise of the same
level.

\section{Conclusions}

We have presented a class of operations - the {\it bi-entangling}
operations - that may be efficiently simulated classically, as
they are only capable of generating two party entanglement. In
some situations this class of operations may give tighter bounds
than currently known on the classical noise tolerance of quantum
gates. One example is the case of depolarizing noise on the CNOT,
for which we show that 67\% noise is sufficient to make the
subsequent evolution efficiently tractable classically, compared
to the best previous bound of 74\%. Another extreme case is with
measurement based computation, where we observe that two-qubit
non-degenerate measurements cannot enable exponential speedup over
classical computation. It may be difficult to extend the class of
bi-entangling operations and still generate a class that is
efficiently tractable. This is because any natural generalizations
to higher numbers of input particles enable perfect quantum
computation, and any extensions that still involve two-particle
gates are hampered by the subtle interplay between separability
preserving and separable gates.

In the second half of this work we turn to bounds on classical
tolerance that may be derived from the Gottesman-Knill theorem.
The subsequent bounds (e.g. 30\% depolarizing noise on the $\pi/8$
gate, 15\% for general single-gate noise) can be relatively low
for this kind of approach.

In general it is quite likely that the bounds derived here may be
improved. One interesting possibility is that a hybrid of the
approaches used by \cite{AB} and \cite{HN} may be used to
understand when slightly non-separable gates may be efficiently
simulated classically, albeit with a noise model more in the
spirit of \cite{AB,R}, where noise is applied to every qubit at
every time step.

In terms of the Clifford gate based work, it seems quite possible
that if the recent conjecture of Bravyi \& Kitaev \cite{BK} is
true, then the fault tolerant threshold gates of the form
$U(\theta) = |0\rangle\langle 0| + \exp(i \theta) |1\rangle\langle
1|$ can indeed be made fault tolerant to the noise levels derived
here (and implied by their work). Their conjecture implies that a
supply of single qubit quantum states from outside the octahedron
$O$ may be `purified' to certain `magic' pure states by the use of
Clifford operations only. As Clifford operations may be made fault
tolerant to some degree via encoding schemes based on Clifford
operations only (see e.g. \cite{NC} and references therein), it
may be possible that the 15\% noise level on the $\pi/8$ gate may
indeed be tolerated as long as the remaining Clifford operations
act within their own (potentially much tighter) fault tolerant
threshold.

These results show the potential of analyzing classical
tractability with the aim of bounding from above fault tolerance
thresholds. Moreover, investigating the quantum/classical
computational transition for (noisy) quantum evolution is
important in its own right \cite{AB,BK}, particularly as there is
the possibility of `intermediate' quantum computation. It may well
be the case that noisy quantum devices cannot be simulated
efficiently classically, yet cannot be used for fault tolerant
quantum computation. This would imply the existence of an
intermediate physical device - such as a noisy quantum system
controlled by a universal classical computer - which is clearly
universal for computation, is better-than-classical as it can
simulate itself efficiently, and yet is not as powerful as a full
quantum computer. Such intermediate devices may be easier to
construct, and hence may provide a more achievable experimental
target.

\bigskip

\section{Acknowledgements}

We thank Koenraad Audenaert and Terry Rudolph for interesting
discussions. We acknowledge financial support by US-Army through
grant DAAD 19-02-0161, The Nuffield Foundation, a Royal Society
Leverhulme Trust Senior Research Fellowship and the EU Thematic
Network QUPRODIS.


\begin{references}
\bibitem{foot} In fact the quantum operation conjectured by
\cite{HN} as an example of an SP operation that is not in the
convex hull of separable and separable+swap operations is in fact
correct. The proof works by noting that LOCC operations on
separable and separable+swap Jamiolkowski states cannot result in
entanglement between particles A1,A2. However, the SP operation
presented in \cite{HN} has this property.
\bibitem{foot2} The symmetry group (\ref{commutant}) is local across
the (A1A2)-(B1B2) split for all unitaries, and so group averaging
does not change the EB properties of any EB state. Hence in
looking for the minimal noise that takes a given unitary into an
EB operation, the form of the noise may be restricted to being
symmetric. As any state that is averaged under (\ref{commutant})
is automatically a valid Jamiolkowski state, may immediately
utilise the results on robustness of entanglement for density
matrices, and we find that all unitaries are equally robust to
being taken to an EB channel by generic noise - this is
unsurprising as all unitaries have Jamiolkowski states that are
maximally entangled across the EB split. We find that a minimum of
3/4 = 75\% noise is required to take a two-qubit unitary into an
entanglement breaking channel.
\bibitem{same} The term `same noise' does not have a strong
meaning here, as whether one form of noise on one gate is the
`same' as a form of noise on another gate strongly depends upon
precise physics of real devices. For example, the same physical
source of noise, such as a fluctuating magnetic field, may affect
different gates in very different ways.
\bibitem{dep} Of course to do this the noise itself must not be a
Clifford operation, so for example dephasing cannot display this
phenomenon.
\end{references}
\end{document}